\documentclass[aps,prc,showpacs,a4paper,twocolumn,nofootinbib]{revtex4}
\usepackage[dvipdfmx]{graphicx}
\usepackage{graphicx}
\usepackage{dcolumn}
\usepackage{bm}
\usepackage{color}
\usepackage{ulem}

\begin{document}

\title{Why does the sign problem occur in evaluating the overlap of HFB wave functions?}

\author{Takahiro Mizusaki$^1$, Makito Oi$^1$ and Noritaka Shimizu$^{2}$}
\affiliation{$^1$Institute of Natural Sciences, Senshu University,
3-8-1 Kanda-Jinbocho, Chiyoda-ku,Tokyo 101-8425, Japan\\
$^2$Center for Nuclear Study, The University of Tokyo, 7-3-1 Hongo,
Bunkyo-ku, Tokyo 113-0033, Japan}


\begin{abstract}
For the overlap matrix element between Hartree-Fock-Bogoliubov states,
there are two analytically different formulae: one with the square root of the determinant (the Onishi formula) and the other with the Pfaffian (Robledo's Pfaffian formula).
The former formula is two-valued as a complex function, hence it leaves the sign of the norm overlap undetermined (i.e., the so-called sign problem of the Onishi formula). On the other hand, the latter formula does not suffer from the sign problem. The derivations for these two formulae are so different that the reasons are obscured why the resultant formulae possess different analytical properties. In this paper, we discuss the reason why the difference occurs by means of the consistent framework, which is based on the linked cluster theorem and the product-sum identity for the Pfaffian. Through this discussion, we elucidate the source of the sign problem in the Onishi formula.
We also point out that different summation methods of series expansions may result in analytically different formulae.
\end{abstract}

\maketitle

\section{Introduction}

The Hartree-Fock-Bogoliubov (HFB) theory gives a simple but profound basis for the nuclear many-body problem where the competition between the nuclear pairing and deformation plays a primary role in the determination of the ground state as well as the excited states. Especially, a combination of the HFB method with the technique of angular momentum projection allows the direct comparison between the theoretical calculations and the experimental data. The projected HFB states can produce more elaborate and accurate calculations although the simplicity of the HFB wave functions is kept from a mean-field point of view. In this way, not only the simple HFB state but also a superposition of different HFB states ($i.e.$, the projected HFB state) have been extensively used for nuclear-structure studies.
Behind this success of the HFB theory, there was a hidden problem for the overlap matrix element between the HFB states. 

Half a century ago, a formula for the overlap matrix element was derived by Onishi and Yoshida \cite{OY66} and is called the Onishi formula \cite{RS80}. 
To derive the Onishi formula, we begin with the Thouless representation 
 \cite{RS80},\cite{Th60} of the HFB wave functions, $ | \phi^{(k)} \rangle$ ($k=0, 1 $)  defined as
\begin{equation}
 | \phi^{(k)} \rangle= e^{\frac{1}{2}\sum_{p,q=1}^N
  Z^{k}_{p,q}c^{\dagger}_p c^{\dagger}_q   } |-\rangle ,
 \label{hfbwave}
\end{equation}
where $c^{\dagger}$'s are the creation operators and $ |-\rangle $ is the bare Fermion vacuum with  
$c_i |-\rangle =0$ ($i=1,\cdots,N$).
The dimension of the Fermion single-particle space is $N$. $Z$ is an $N\times N$ complex skew-symmetric matrix.
The Thouless representation is a specific one of the Bogoliubov quasiparticle states. 
In this representation, the overall phase is fixed for two HFB wave functions $ | \phi^{(0)} \rangle$ and $ | \phi^{(1)} \rangle$, respectively, as in Refs.\cite{NW83},\cite{Rob09}.  

The overlap matrix element between these two HFB wave functions is defined as 
\begin{equation}
  \langle \phi^{(0)} | \phi^{(1)} \rangle =
 \langle - |  
e^{\frac{1}{2}\sum_{p',q'=1}^N Z^{0*}_{p',q'} c_{q'} c_{p'}   }
 e^{\frac{1}{2}\sum_{p,q=1}^N Z^{1}_{p,q} c^{\dagger}_p c^{\dagger}_q   } |-\rangle,
\label{overlapzz}
\end{equation}
which can be expressed as 
\begin{equation}
 \langle \phi^{(0)} | \phi^{(1)} \rangle= \sqrt{ \textrm{ Det} (I-Z^{0*} Z^{1}  )}.
 \label{onishi_formula}
\end{equation}
This formula is known as the Onishi formula \cite{OY66}.
Due to the square root function, the Onishi formula is two-valued and does not give a definite sign if $Z$'s are complex matrices. This indefiniteness of the sign assignment is referred to as the sign problem of the Onishi formula, which becomes quite serious in the application of the full angular momentum projection.
So far, there are several approaches known to remedy the problem \cite{NW83}, \cite{Rob09}, \cite{HHR82},  \cite{OT05},  \cite{BD17}.

Among them, Robledo  \cite{Rob09} has recently derived an alternative and ambiguity-free formula for the overlap matrix element by the Pfaffian as
\begin{equation}
 \langle \phi^{(0)} | \phi^{(1)} \rangle= s_N {\it Pf} \left[\begin{array}{cc}
     Z^{1} & -I \\
      I &  -Z^{0*}  \\
    \end{array}\right] ,
\label{robledo_pfaffian}
\end{equation}
where $s_N =(-)^{N(N+1)/2}$ and $I$ is the $N \times N$ identity matrix.
This formula is proved with rather advanced techniques, that is, the Fermion coherent state and Grassmann integral. 
His proof is mathematically very elegant and interesting \cite{Rob09}. Moreover, these techniques led us to a relation to the generalized Wick's theorem and its related topics \cite{OM11}, \cite{BR12}, \cite{AB12}, \cite{MO12}, \cite{MO13}. The proof by Robledo is, however, rather abstract and 
it somewhat keeps us from an intuitive understanding of the reason for the disappearance of the sign ambiguity.

In the present paper, we derive both formulae in Eqs. (\ref{onishi_formula}) and (\ref{robledo_pfaffian}) directly from Eq.(\ref{overlapzz}) and elucidate an origin of the sign problem. 
First, we expand the exponential operators in Eqs.(\ref{hfbwave},\ref{overlapzz}). 
After handling the vacuum expectation values of the product of the creation-annihilation operators, the overlap matrix element can be, in principle, expressed as a polynomial of the matrix elements of $Z$.  The overlap is, therefore,  single-valued.

Next, we will consider two summation methods. 
We show that an expansion of the HFB wave function in Eq.(\ref{hfbwave}) can be expressed by
the Pfaffians and that the overlap matrix element can be, thereby, revealed by a finite series of the product-sum of the Pfaffians. This finite series can be summed up into Robledo's Pfaffian formula in Eq.(\ref{robledo_pfaffian}).
By this derivation, Robledo's Pfaffian formula is turned out to be obviously single-valued and to be free of the sign problem.
We also show that the other summation method with the linked cluster theorem\cite{BM63} brings us to the Onishi formula. 
We  present that this summation concerning the connected diagrams involves an infinite series, which is in sharp contrast to the original finite series and that it gives rise to the square root function in the Onishi formula.  
We also clarify that the skew-symmetric property of the Thouless matrix $Z$ can remove the sign problem from the Onishi formula completely.

The present paper is organized as follows.
In Sec. II, we show a basic structure of the overlap 
matrix element through the series expansion of the HFB wave functions and present an alternative derivation of Robledo's Pfaffian formula.
In Sec. III, we show a relation between the Onishi formula and the linked cluster theorem, and we discuss the origin of the sign problem.  
In Sec. IV, we give a conclusion. In the appendices, we summarize useful identities concerning the Pfaffian and show the derivation of the connected term.

\section{Overlap formula with the Pfaffian}
\subsection{Basic structure of the overlap matrix element}

First, we show a basic mathematical structure of the overlap matrix element by expanding the
exponential operators in Eq.(\ref{overlapzz}).
Defining pair-annihilation and pair-creation operators, $\hat{A}$ and $\hat{B}$ as 
\begin{eqnarray}
\hat{A}&=&  \frac{1}{2}\sum_{p,q} Z^{0*}_{p,q} c_q c_p=\sum_{q>p} Z^{0*}_{p,q} c_q c_p  , \\
\hat{B}&=&  \frac{1}{2}\sum_{p,q} Z^{1}_{p,q} c^{\dagger}_p c^{\dagger}_q=\sum_{q>p}Z^{1}_{p,q} c^{\dagger}_p c^{\dagger}_q,
\end{eqnarray}
the HFB wave functions are shown by 
\begin{eqnarray}
\langle \phi^{(0)} | &=& \langle -  | e^{\hat{A}} ,  \nonumber \\
| \phi^{(1)} \rangle &=& e^{\hat{B}} |- \rangle .
\label{defAB}
\end{eqnarray}
The overlap matrix element is simply denoted by
\begin{equation}
  \langle \phi^{(0)} | \phi^{(1)} \rangle =
 \langle - |  
e^{\hat{A}} e^{\hat{B}}|-\rangle.
\label{overlapAB}
\end{equation}

By expanding exponential operator, the HFB wave function can be shown as  
\begin{equation}
| \phi^{(1)} \rangle = ( 1+\hat{B}+\frac{1}{2!}\hat{B}^2 +\frac{1}{3!}\hat{B}^3+ \cdots +\frac{1}{ \left( \frac{N}{2} \right) !}\hat{B}^{\frac{N}{2}} ) |- \rangle , 
\end{equation}
where this series expansion terminates in order $N/2$ because the number of single particle states, namely, the dimension of the matrices $Z^0$ and $Z^1$ is $N$.
The overlap matrix element is rewritten by 
\begin{equation}
 \langle \phi^{(0)} | \phi^{(1)} \rangle 
  =\sum_{k=0}^{N/2} \langle -| \frac{\hat{A}^k}{k!} \frac{\hat{B}^k}{k!} |- \rangle.
\label{AB_k}
\end{equation}
Note that  $\langle -| \frac{\hat{A}^l}{l!} \frac{\hat{B}^k}{k!} |- \rangle$ vanishes if $l \neq k$
because $\hat{A}$ and $\hat{B}$ are pair-annihilation and pair-creation operators, respectively.

Next, we investigate an expanded form of the $\frac{1}{k!}\hat{B}^k$ operator.
The $\frac{1}{k!}\hat{B}^k$ operator is generally expressed by the $2k$ creation operators with the coefficients in terms of matrix elements of $Z^{1}$ as, 
\begin{equation}
\frac{1}{k!} \hat{B}^k 
= \frac{1}{k!} \frac{1}{2^k} \sum  Z^{1}_{p_1,q_1} \cdots {Z^{1}}_{p_k,q_k}  c^{\dagger}_{p_1} c^{\dagger}_{q_1} \cdots  c^{\dagger}_{p_k} c^{\dagger}_{q_k}.
\label{Bk0}  
\end{equation}
As the $\hat{A}^k$ operator is also similarly shown, the overlap matrix element can be straightforwardly expressed as 
a function of matrix elements of $Z^{0*}$ and $Z^{1}$ as,
\begin{eqnarray}
  & &\langle \phi^{(0)} | \phi^{(1)} \rangle  \nonumber \\
  &=&  \sum_{k=0}^{N/2} \frac{1}{\left(k!\right)^2}  \frac{1}{2^{2k}} \sum_{p,q,p',q'}Z^{0*}_{p'_1,q'_1} \cdots {Z^{0*}}_{p'_k,q'_k}  Z^{1}_{p_1,q_1} \cdots {Z^{1}}_{p_k,q_k} \nonumber \\
  & & \langle - | c_{q'_k}  c_{p'_k} \cdots c_{q'_1}  c_{p'_1}  
   c^{\dagger}_{p_1} c^{\dagger}_{q_1} \cdots  c^{\dagger}_{p_k} c^{\dagger}_{q_k} |- \rangle.
 \label{AkBk_structure}  
\end{eqnarray}
The matrix element in the third line of the above equation gives intricate restriction concerning $p$'s, $q$'s,
$p'$'s, $q'$'s by taking the contractions.
The above formula shows a very complicated structure regarding the matrix elements of $Z^{0*}$ and $Z^{1}$. It is, however, quite evident that the overlap matrix element is a polynomial of the matrix elements of $Z^{0*}$ and $Z^{1}$ and has, thereby, no sign ambiguity. 

In the subsequent subsections, we will show that Eq.(\ref{AkBk_structure}) can be rewritten by the Pfaffians and will directly derive Robledo's Pfaffian formula from Eq.(\ref{AkBk_structure}).
Furthermore, in the next section, by handling Eq.(\ref{AkBk_structure}) with the linked cluster theorem, we will derive the Onishi formula.  

\subsection{Overlap formula with product-sum of the Pfaffians}

In this subsection, we consider to rewrite Eq.(\ref{AkBk_structure}) by investigating the detailed structure of Eq.(\ref{Bk0}). 

For example, the 2nd order term is expressed as
\begin{eqnarray}
\frac{1}{2!} \hat{B}^2  &=& \frac{1}{2!} \sum_{p_1<q_1,p_2<q_2}Z^{1}_{p_1, q_1} Z^{1}_{p_2, q_2} c^{\dagger}_{p_1} c^{\dagger}_{q_1} c^{\dagger}_{p_2} c^{\dagger}_{q_2}  , \nonumber \\
                   &=& \sum_{p_1<q_1,p_2<q_2,p_1<p_2}Z^{1}_{p_1, q_1} Z^{1}_{p_2, q_2}  c^{\dagger}_{p_1} c^{\dagger}_{q_1} c^{\dagger}_{p_2} c^{\dagger}_{q_2},  \nonumber \\ 
\label{B2}                   
\end{eqnarray}
where $2!$ is removed due to the additional condition $p_1 < p_2$. 
Now let us change the integer indices  $p_1$, $q_1$, $p_2$, $q_2$ to new indices $n_1$, $n_2$, $n_3$, $n_4$ with $n_1 < n_2 < n_3 < n_4$, and let us obtain the coefficients $b_{\{n\}}$ in the following form;
\begin{equation}
\frac{1}{2!} \hat{B}^2  
= \sum_{n_1 < n_2 < n_3 < n_4}b_{\{n\}}c^{\dagger}_{n_1} c^{\dagger}_{n_2} c^{\dagger}_{n_3}c^{\dagger}_{n_4} 
\end{equation} 
where $n$'s run $1$ to  $N$ under the condition $n_1 < n_2 < n_3 < n_4$.
These two kinds of indices have different conditions.
The relation between them is classified into the following three cases:
(1) $p_1=n_1$, $q_1=n_2$, $p_2=n_3$, $q_2=n_4$, and $c^{\dagger}_{p_1} c^{\dagger}_{q_1} c^{\dagger}_{p_2}c^{\dagger}_{q_2}=c^{\dagger}_{n_1} c^{\dagger}_{n_2} c^{\dagger}_{n_3}c^{\dagger}_{n_4}$,
(2) $p_1=n_1$, $q_1=n_3$, $p_2=n_2$, $q_2=n_4$,  and   $c^{\dagger}_{p_1} c^{\dagger}_{q_1} c^{\dagger}_{p_2}c^{\dagger}_{q_2}=-c^{\dagger}_{n_1} c^{\dagger}_{n_2} c^{\dagger}_{n_3}c^{\dagger}_{n_4}$,
(3)  $p_1=n_1$, $q_1=n_4$, $p_2=n_2$, $q_2=n_3$, and $c^{\dagger}_{p_1} c^{\dagger}_{q_1} c^{\dagger}_{p_2}c^{\dagger}_{q_2}=c^{\dagger}_{n_1} c^{\dagger}_{n_2} c^{\dagger}_{n_3}c^{\dagger}_{n_4}$.
Therefore, the coefficients $b_{\{n\}}$ can be given in terms of the Pfaffian as
\begin{eqnarray}
b_{\{n\}}
&=& Z^{1}_{n_1,n_2} Z^{1}_{n_3,n_4} -Z^{1}_{n_1,n_3} Z^{1}_{n_2,n_4} 
+Z^{1}_{n_1,n_4} Z^{1}_{n_2,n_3},  \nonumber \\
&=& {\it Pf}\left[\begin{array}{cccc}
      0            & Z^{1}_{n_1,n_2} & Z^{1}_{n_1,n_3} & Z^{1}_{n_1,n_4} \\
-Z^{1}_{n_1,n_2} &  0                & Z^{1}_{n_2,n_3} & Z^{1}_{n_2,n_4}  \\
-Z^{1}_{n_1,n_3} & -Z^{1}_{n_2,n_3}& 0                 & Z^{1}_{n_3,n_4}  \\
-Z^{1}_{n_1,n_4} & -Z^{1}_{n_2,n_4}& -Z^{1}_{n_3,n_4}&      0            
        \end{array}\right].
\end{eqnarray} 
where we use Eq.(\ref{pfaff4}).
 
In general, the $\frac{1}{k!}\hat{B}^k$ operator is expressed by the $2k$ creation operators with the coefficients in terms of matrix elements of $Z^{1}$ as, 
\begin{equation}
\frac{1}{k!} \hat{B}^k 
= \sum{Z^{1}}_{p_1,q_1} \cdots {Z^{1}}_{p_k,q_k}  c^{\dagger}_{p_1} c^{\dagger}_{q_1} \cdots  c^{\dagger}_{p_k} c^{\dagger}_{q_k},
\label{Bk_structure}  
\end{equation}
where the summations are performed with the restriction, $p_1 < q_1 , \cdots, p_k < q_k , p_1 < \cdots < p_k$. The $k!$ is removed due to the additional condition $p_1< \cdots < p_k$. 
As the same procedure, we change the integer indices $p_1, q_1, \cdots, p_k, q_k$ 
($p_1 < q_1 , \cdots, p_k < q_k , p_1 < \cdots < p_k$)
to $2k$ different integer indices  $n_1, \cdots, n_{2k}$ ($n_1 < \cdots < n_{2k}$). 
As this condition is the same as Eqs.(\ref{pfcond},\ref{defpfaff1}), we can introduce the Pfaffian as 
\begin{equation}
\frac{1}{k!} \hat{B}^k 
 = \sum_{n_1 < \cdots < n_{2k}} {\it Pf}\left[\mathcal{Z}^{1}_{2k}\right]  c^{\dagger}_{n_1} \cdots c^{\dagger}_{n_{2k}},
\end{equation}
where $n$'s run $1$ to $N$ under the restriction, $n_1 < \cdots < n_{2k}$ and the $m \times m$ skew-symmetric matrix $\mathcal{Z}^{1}_{m}$ is defined as
\begin{equation}
\mathcal{Z}^{1}_{m}=
\left[\begin{array}{ccccc}
                 0 & Z^{1}_{n_1,n_2}  &         &  \cdots                 & Z^{1}_{n_1,n_{m}} \\
-Z^{1}_{n_1,n_2} &  0                 &         &  \cdots                 &      \vdots       \\
     \vdots        &                    &  \ddots &                &      \vdots        \\
     \vdots        &       \cdots       &         &     0          & Z^{1}_{n_{m-1},n_{m}}  \\
-Z^{1}_{n_1,n_{m}} &       \cdots       &         & -Z^{1}_{n_{m-1},n_{m}} &      0            
    \end{array}\right].
\end{equation}
The HFB wave functions are expressed as 
\begin{eqnarray}
| \phi^{(i)} \rangle &=& \Bigl[
1+\sum_{n_1<n_2} {\it Pf} \left[ \mathcal{Z}^{i}_2 \right]  c^{\dagger}_{n_1} c^{\dagger}_{n_2} +  \nonumber  \\  
&&\sum_{n_1 < \cdots< n_4} {\it Pf} \left[\mathcal{Z}^{i}_4 \right]  c^{\dagger}_{n_1} c^{\dagger}
_{n_2}c^{\dagger}_{n_3} c^{\dagger}_{n_4}
+ \cdots +  \nonumber \\ 
&&\sum_{n_1 < \cdots < n_N} {\it Pf} \left[\mathcal{Z}^{i}_N \right]  c^{\dagger}_{n_1} \cdots c^{\dagger}_{n_N} \Bigl] |- \rangle,  
\label{HFBwavePf}
\end{eqnarray}
where $i$ takes 0 or 1. The expansion of the HFB wave function can be generally shown 
by the Pfaffians. 

The overlap matrix element, $\langle \phi^{(0)}|\phi^{(1)} \rangle $, is given in terms of the Pfaffians as
\begin{eqnarray}
\langle \phi^{(0)}|\phi^{(1)} \rangle  &=&  1+\sum_{n_1<n_2} {\it Pf}\left[\mathcal{Z}^{0*}_2\right] {\it Pf}\left[\mathcal{Z}^{1}_2\right]
\nonumber \\
&+&  \sum_{n_1 < \cdots < n_4} {\it Pf}\left[\mathcal{Z}^{0*}_4\right] {\it Pf}\left[\mathcal{Z}^{1}_4\right] + \cdots  \nonumber \\
&+&\sum_{n_1 < \cdots < n_N} {\it Pf}\left[\mathcal{Z}^{0*}_N\right]{\it Pf}\left[\mathcal{Z}^{1}_N\right].
\end{eqnarray}
By introducing an index set ${\mathcal I}= \{n_1,n_2, \cdots, n_{2t}\}$ ($n_1 < \cdots < n_{2t}$), we can define the sub-matrix $\mathcal{Z}^{1}_{2t}$
as $Z^{1}_{\mathcal I}$, which is defined as
\begin{equation}
 (Z^{1}_{\mathcal I})_{i,j}= Z^{1}_{n_i, n_j},
\end{equation}
where $i,j$ run $1$ to $2t$.
With this notation, the overlap matrix element is compactly expressed as 
\begin{equation}
\langle \phi^{(0)}|\phi^{(1)} \rangle  = \sum_{t=0}^{N/2} \sum_{{\mathcal I} \in I^{N}_{2t}} 
{\it Pf}\left[ Z^{0*}_{\mathcal I}\right ] Pf\left[Z^{1}_{\mathcal I}\right] ,
\label{overlapPfPf}
\end{equation}
where $I^{N}_{2t}$ is a set, consisting of subsets with $2t$ elements in $\{ 1,2,3, \cdots, N \}$. 
For example, for $N=4$, the set $\mathcal I$ takes $\{ \}$, $\{1,2\}$,
$\{1,3\}$, $\{2,3\}$, $\{1,4\}$, $\{2,4\}$, $\{3,4\}$ and $\{1,2,3,4\}$.
In general, the summation is taken over $2^{N-1}$ elements of $I^{N}_{2t}$.
Therefore, in numerical computation, this overlap formula with the product-sum of the Pfaffians, Eq.(\ref{overlapPfPf}), is not practical if $N$ increases. 

\subsection{Derivation of Robledo's Pfaffian formula without the Grassmann integrals}
Next, we show that the Pfaffians in Eq.(\ref{overlapPfPf}) can be expressed with single Pfaffian, thanks to the product-sum identity of the Pfaffian, Eq.(\ref{prosum}) in Appendix B.
We define two skew-symmetric matrices $P$ and $Q$ as
\begin{eqnarray}
P&=&\left[\begin{array}{cc}
     {Z^{1}} &    0      \\
        0    &  -{Z^{0*}}  
     \end{array}\right   ],      
\nonumber  \\ 
Q&=&\left[\begin{array}{cc}
        0    &    -I      \\
        I    &     0    
     \end{array}\right   ].    
\label{defPQ}       
\end{eqnarray}
The l.h.s. of Eq. (\ref{prosum}) is rewritten as 
\begin{equation}
{\it Pf} \left[P+Q \right] ={\it Pf} \left[\begin{array}{cc}
                {Z^{1}}    &    -I      \\
                I    &    -{Z^{0*}}    
             \end{array}\right   ],      
\end{equation}
which is Robledo's Pfaffian, except $s_N$. 
As the dimension of ${Z^{0}}$ and ${Z^{1}}$ is $N$,
$m=2N$ in Eq.(\ref{prosum}).
The r.h.s. of Eq.(\ref{prosum}) becomes 
\begin{equation}
{\it Pf}\left[P+Q \right]=\sum_{r=0}^{N}\sum_{ \mathcal{I} \in I^{2N}_{2r}}(-1)^{|\mathcal{I} |-r}
{\it Pf}\left[P_{\mathcal{I} }\right]{\it Pf}\left[Q_{\overline{\mathcal{I}}}\right],
\label{PQsum}
\end{equation}
where $I^{2N}_{2r}$ is a set, consisting of subsets with $2r$ elements in $\{ 1,2, \cdots, 2N \}$.
Below, we will show that Eq.(\ref{PQsum}) can be reduced to Eq.(\ref{overlapPfPf}) by the Pfaffian identities and classification of the indices in Eq.(\ref{PQsum}).

As the matrices $P$ and $Q$ have a bipartite structure, we divide the  $\mathcal{I}$ into two parts $\mathcal{I}_1$ and $\mathcal{I}_0$, $\mathcal{I}=\mathcal{I}_1 \bigoplus \mathcal{I}_0$, where
 $\mathcal{I}_1$  is  a subset of  $\{ 1,2, \cdots, N \}$ and   
    $\mathcal{I}_0$  is  a subset of  $\{ N+1,N+2, \cdots, 2N \}$. 
For example, for $N=4$ and $r=2$, elements of the set $I^{2N}_{2r}=I^8_4$ are
$\{ 1,2,5,6\}$, $\{ 1,3,7,8\}$,  $\{ 1,2,4,5\}$ and so on. For the set $\{ 1,2,5,6\}$, 
$\mathcal{I}_1$ is $\{ 1,2\}$ and $\mathcal{I}_0$ is $\{ 5,6\}$.  
For the set $\{ 1,3,7,8\}$, 
$\mathcal{I}_1$ is $\{ 1,3\}$ and $\mathcal{I}_0$ is $\{ 7,8\}$. The numbers of elements of both sets are the same.  
For the set $\{ 1,2,4,5\}$, 
$\mathcal{I}_1$ is $\{ 1,2,4\}$ and $\mathcal{I}_0$ is $\{ 5\}$. The numbers of elements of both sets are different. 

We classify the $\mathcal{I}$ by the symmetry with the bipartite structure. 
One is symmetric concerning $\mathcal{I}_1$ and $\mathcal{I}_0$ and is denoted as $\mathcal{I}^{s}$.
The $\mathcal{I}^{s}$ with numbers of elements $2r$ is a sum of $\mathcal{I}_1=\{l_1,l_2, \cdots, l_r\}$ 
and $\mathcal{I}_0=\{l_1+N,l_2+N, \cdots, l_r+N\}$.
The $\{ 1,2,5,6\}$ is an example of $\mathcal{I}^{s}$.
The other is asymmetric and is called $\mathcal{I}^{a}$, whose examples are $\{ 1,3,7,8\}$, $\{ 2,4,5,8\}$.
Only these symmetric sets $\mathcal{I}^{s}$ with even integer $r$ can contribute to Eq.(\ref{PQsum}), while other sets, namely, symmetric sets $\mathcal{I}^{s}$ with odd integer $r$ and asymmetric sets $\mathcal{I}^{a}$  give no contribution.

In the former case, ${\mathcal I}^{s}$ with $2r$ ($r$:odd), matrix dimensions of $Z^{1}_{{\mathcal I}^{s}_1} $ and $Z^{0}_{{\mathcal I}^{s}_1} $ are odd. For example, we consider a symmetric case of $N=4$ and $r=3$. Elements of the set $I^{2N}_{2r}=I^8_6$ are $\{ 1,2,3,5,6,7\}$, $\{ 1,2,4,5,6,8\}$, and so on. The numbers of elements of ${\mathcal I}^{s}_1$ are 3.
The Pfaffians of $Z^{1}_{{\mathcal I}^{s}_1} $ and $Z^{0}_{{\mathcal I}^{s}_1} $ are, thereby, zero.
The Pfaffian of  $P_{{\mathcal I}^{s} }$ becomes 
\begin{eqnarray}
{\it Pf}\left[P_{{\mathcal I}^{s} }\right   ]&=&
{\it Pf} \left[\begin{array}{cc}
                {Z^{1}}    &    0     \\
                0    &    -{Z^{0*}}    
             \end{array}\right   ]_{{\mathcal I}^{s}}\nonumber \\
             &=&{\it Pf} \left[ { Z^{1}}_{{\mathcal I}^{s}_1}  \right]
              {\it Pf} \left[ {-Z^{0*}}_{{\mathcal I}^{s}_1}  \right] =0,   
\end{eqnarray}
where we use Eq.(\ref{pf12}). Therefore, the number of elements of ${\mathcal I}^{s}$, giving non-vanishing Pfaffian, is $4t$ ($t$:integer).

In the latter case, if the numbers of elements of ${\mathcal I}^{a}_{1}$ and 
${\mathcal I}^{a}_{0}$ are different,  ${\it  Pf}(Q_{\overline{{\mathcal I}^{a}}})$=0 because
$Q_{\overline{{\mathcal I}^{a}}}$ is not a square matrix. 
For example, we again take the case of  $\{ 1,2,4,5\}$ for $N=4$ and $r=2$. The numbers of elements of  ${\mathcal I}^{a}_{1}$ and ${\mathcal I}^{a}_{0}$ are 3 and 1, respectively. The matrix $Q_{\overline{{\mathcal I}^{a}}}$ is $1 \times 3$ and is not a square one.
If the numbers of elements of ${\mathcal I}^{a}_{1}$ and ${\mathcal I}^{a}_{0}$ are the same and are $2t$, $Pf(Q_{\overline{{\mathcal I}^{a}}})$=0 due to asymmetry of index. 
For instance, we again take the case of  $\{ 1,3,7,8\}$ for $N=4$ and $r=2$. 
Its complementary set is $\{2,4,5,6 \}$. 
The $Q_{\overline{{\mathcal I}^{a}}}$ is given by
\begin{equation}
\left[\begin{array}{cccc}
        q_{2,2}  &  q_{2,4} & q_{2,5} & q_{2,6}   \\
        q_{4,2}  &  q_{4,4} & q_{4,5} & q_{4,6}   \\
        q_{5,2}  &  q_{5,4} & q_{5,5} & q_{5,6}   \\
        q_{6,2}  &  q_{6,4} & q_{6,5} & q_{6,6} 
     \end{array}\right ]=
\left[\begin{array}{cccc}
        0  & 0 & 0 & -1   \\
        0  & 0 & 0 & 0   \\
        0  & 0 & 0 & 0   \\
        1  & 0 & 0 & 0 
     \end{array}\right ]       
\end{equation} 
because of the definition of $Q$, Eq.(\ref{defPQ}). 
Therefore, its Pfaffian is zero. 
In general, $Q_{\overline{{\mathcal I}^{a}}}$ is shown in the bipartite form as
\begin{equation}
{\it Pf}\left[ Q_{\overline{{\mathcal I}^{a}}}\right ] ={\it Pf}
 \left[ \begin{array}{cc}
                0    &    C      \\
               -C^T   &      0         
             \end{array}\right ],
\label{Q_bar_index}  
\end{equation} 
where the diagonal block matrices are zero and the off-diagonal block matrix is denoted by $C$.
By the identity Eq.(\ref{CCpf}), 
it reduces to $(-1)^{m(m-1)/2}\textrm{ Det} \left[  C\right ]$ where $m=N-2t$.
Let $\overline{{\mathcal I}^{a}_1}$ and $\overline{{\mathcal I}^{a}_0}$ be $\{ i_1, i_2,\cdots, i_m \}$
$(i_1 < i_2 < \cdots < i_m)$ and $\{ j_1+N, j_2+N,\cdots, j_m+N \}$ $(j_1 < j_2 < \cdots < j_m)$, respectively. As some indices are the same and others are different, we re-sort the indices as
$\{ i_1, i_2,\cdots, i_m \} \rightarrow \{ {i'}_1, \cdots,{i'}_k,\cdots, {i'}_m \} $ and 
$\{ j_1, j_2,\cdots, j_m \} \rightarrow \{ {j'}_1, \cdots,{j'}_k,\cdots, {j'}_m \} $ where
${i'}_i={j'}_i $ for ($ i \le k$) and ${i'}_i \ne {j'}_i $ for ($ i > k$). 
In this representation, the matrix $C$ can be shown as  
\begin{equation}
C=
\left[\begin{array}{cccccc}
       -1  &        &    &   &        &     \\
           & \ddots &    &   &        &     \\
           &        & -1 &   &        &     \\
           &        &    & 0 &        &     \\
           &        &    &   & \ddots &     \\
           &        &    &   &        &  0  \\
     \end{array}\right ],     
\end{equation} 
while the diagonal block matrices in Eq.(\ref{Q_bar_index}) are still zero.
Therefore, $Pf(Q_{\overline{{\mathcal I}^{a}}})=0$ is proved.

Next we consider  ${\mathcal I}^{s}$  with $4t$ ($t$:integer), which consists of 
${{\mathcal I}_1}^{s}=\{l_1,l_2, \cdots, l_{2t}\} $ and
${{\mathcal I}_0}^{s}=\{l_1+N,l_2+N, \cdots, l_{2t}+N\} $. 
For instance, such a case is $\{ 1,2,5,6\}$ for $N=4$ and $t=1 (r=2)$. 
The Pfaffian of $P$ is rewritten as  
\begin{eqnarray}
{\it Pf}\left[\begin{array}{cc}
     {Z^{1}} &    0      \\
        0    &  -{Z^{0*}}  
     \end{array}\right ]_{{\mathcal I}^{s}}  
   &=& {\it Pf}\left[{Z^{1}}_{{\mathcal I}^{s}_1} \right]Pf\left[-{Z^{0*}}_{{\mathcal I}^{s}_1} \right]  \nonumber \\
   &=& {\it Pf}\left[{Z^{1}}_{{\mathcal I}^{s}_1} \right]Pf\left[{Z^{0*}}_{{\mathcal I}^{s}_1} \right](-1)^{t},\nonumber \\ 
\label{P_bunkai}
\end{eqnarray} 
where $Z^{1}_{{\mathcal I}^{s}_1}$ and $ Z^{0*}_{{\mathcal I}^{s}_1}$ are $2t \times 2t$ matrices, 
and we use Eqs.(\ref{minuspf},\ref{pf12}). 
The Pfaffian of $Q$ is evaluated as 
\begin{eqnarray}
{\it Pf}\left[\begin{array}{cc}
        0 &    -1      \\
        1    &    0   
     \end{array}\right ]_{\overline{{\mathcal I}^{s}}}
   &=& 
{\it Pf}\left[\begin{array}{cc}
        0 &    (-I )_{\overline{{\mathcal I}^{s}_1}}        \\
        I_{\overline{{\mathcal I}^{s}_1}}       &    0   
     \end{array}\right ]   \nonumber \\
   &=& (-1)^{\frac{1}{2}(N-2t)(N-2t-1)} \textit{Det}\left[  (-I )_{\overline{{\mathcal I}^{s}_1}}   \right ]        \nonumber \\
   &=& (-1)^{\frac{1}{2}(N-2t)(N-2t+1)},
\end{eqnarray} 
where $I_{\overline{{\mathcal I}^{s}_1}}$  is a $(N-2t) \times (N-2t)$ identity matrix and we use Eq.(\ref{CCpf}). 
Thus, the product-sum identity of the Pfaffian is reduced into 
\begin{eqnarray}
{\it Pf}  \left[\begin{array}{cc}
                {Z^{1}}    &    -I      \\
                I    &    -{Z^{0*}}    
             \end{array}\right   ]
&=&
\sum_{t=0}^{N/2}\sum_{{\mathcal I}^{s}\in I^{N}_{2t} } (-1)^{|{\mathcal I}^{s}|-2t} 
{\it Pf}\left[Z^{1}_{{\mathcal I}^{s}_{1}}\right] \nonumber \\
&& {\it Pf}\left[Z^{0*}_{{\mathcal I}^{s}_{1}}\right]
 (-1)^{t}  (-1)^{\frac{1}{2}(N-2t)(N-2t+1)}.\nonumber \\
\label{sum_pf_sign}
\end{eqnarray} 
As $|{\mathcal I}^{s}|=2|{\mathcal I}^{s}_1|+2Nt$, $(-1)^{|{\mathcal I}^{s}|}=1$. Thereby, the sign of r.h.s. of Eq.(\ref{sum_pf_sign}) becomes 
$(-1)^{\frac{1}{2}N(N+1)}$,
which is just Robledo's $s_N$ \cite{Rob09}.  Therefore, the obtained overlap matrix element agrees with Robledo's Pfaffian expression as,
\begin{eqnarray}
\langle \phi^{(0)}|\phi^{(1)} \rangle &=&
\sum_{t=0}^{N/2}\sum_{{\mathcal I}_{1}^{s} \in I^{N}_{2t}}  
{\it Pf}\left[Z^{0*}_{{\mathcal I}_{1}^{s}}\right]  
{\it Pf}\left[Z^{1}_{{\mathcal I}_{1}^{s}}\right]  \nonumber \\
&=& s_N Pf  \left[\begin{array}{cc}
                {Z^{1}}    &    -I      \\
                I    &    -{Z^{0*}}    
             \end{array}\right   ].
\end{eqnarray} 
Thus, we algebraically derived Robledo's Pfaffian formula, summing up the expansion terms by applying the product-sum identity of the Pfaffian.
As this expansion forms a finite series and is a polynomial, it is evident that its summation is also single-valued and the obtained formula has no sign problem.                   
                 
\section{Overlap formula with the Determinant}

The Onishi formula was first obtained by Onishi and Yoshida \cite{OY66}, and Onishi and his collaborators used the linked cluster expansion\cite{BM63} for the double-variational method \cite{KO77},\cite{OH80}.
Here, we derive the Onishi formula based on the linked cluster theorem\cite{BM63},
which is more standard in view of the quantum many-body theory.
Especially we focus on the origin of square-root function.

\subsection{Expansion of overlap matrix element with the contractions}
Let us begin with the overlap matrix element in Eq. (\ref{overlapzz}),
which is rewritten as
\begin{equation}
 \langle \phi^{(0)} | \phi^{(1)} \rangle = 
 \langle e^{\hat{A}} e^{\hat{B}}\rangle
  =\sum_{k=0}^{N/2} \langle \frac{\hat{A}^k}{k!} \frac{\hat{B}^k}{k!} \rangle,
\label{AkBk}
\end{equation}
where $N$ is a dimension of the model space. This is the same equation as Eq.(\ref{AB_k}) while here 
we denote a vacuum expectation value simply as 
$\langle\hat{O} \rangle \equiv \langle - |\hat{O} |-\rangle$ where $\hat{O}$ is an arbitrary operator.
As Eq.(\ref{AkBk}) forms a finite series, it seems to be difficult to derive square-root function.
However, by evaluating Eq.(\ref{AkBk}) with contractions, we naturally obtain an infinite series.  
 
Before we discuss the general term, we explicitly show several terms for $k=0 \sim 3$.
The $0$-th order term is unity.
For $k=1$, taking the contractions, we can obtain the 1st order term as, 
\begin{eqnarray}
\langle \hat{A}\hat{B} \rangle 
&=& \frac{1}{2^2} \sum_{p,q,p',q'=1}^{N}Z^{0*}_{p'q'}Z^{1}_{pq} \langle c_{q'} c_{p'} c_p^{\dagger} c_q^{\dagger}   \rangle \nonumber  \\
&=& \frac{1}{2^2} \left[ -2\textrm{Tr}\left( Z^{0*}Z^{1} \right) \right] \nonumber \\
&=& -\frac{1}{2} \textrm{Tr} (Y)
\end{eqnarray} 
where $Y\equiv  Z^{0*}Z^{1}$.
The contribution only from the connected diagrams is denoted
with the suffix "c" attached to the expectation value.
For this notation,  the 1st order term is shown as,
\begin{equation}
\langle \hat{A} \hat{B} \rangle _c=\frac{\textrm{Tr}(Y^1)}{-2\cdot 1},
\end{equation} 
where the 1st power of $Y$ is denoted as $Y^1$ for later convenience. 
The second-order term is shown by 
\begin{eqnarray}
\frac{1}{\left( 2!\right)^2}    \langle  \hat{A}^2 \hat{B}^2  \rangle 
&=&  \frac{1}{\left( 2!\right)^2} \frac{1}{2^4} \sum
Z^{0*}_{p'_1q'_1}Z^{0*}_{p'_2q'_2}Z^{1}_{p_1q_1}Z^{1}_{p_2q_2}  \nonumber \\
& & \langle c_{q'_2} c_{p'_2}c_{q'_1} c_{p'_1} c_{p_1}^{\dagger} c_{q_1}^{\dagger} c_{p_2}^{\dagger} c_{q_2}^{\dagger} \rangle.
\label{AB2expand}
\end{eqnarray}
The contractions are classified into two groups. One is a disconnected term and is a product of 1st-order connected terms. The other is a connected term.
The former one is shown by 
\begin{equation}
\frac{1}{\left( 2!\right)^2}  (2!) \langle \hat{A}\hat{B}  \rangle _c ^2 
=   \frac{1}{2!} \left(\frac{\textrm{Tr}(Y^1)}{-2\cdot 1} \right)^2,
\end{equation}
where $2!$ is the number of the repeated diagrams.
The other is a connected term, which corresponds to
\begin{equation}   
\langle \frac{\hat{A}^2}{2!} \frac{\hat{B}^2}{2!}   \rangle_{c} =\left( \frac{\textrm{Tr}(Y^2)}{-2\cdot 2} \right),
\label{connect2}
\end{equation}
where the coefficient $\frac{1}{-2\cdot 2}$ is explained in Appendix C.
Therefore, the second order term is given by
\begin{equation}
\langle \frac{\hat{A}^2}{2!} \frac{\hat{B}^2}{2!}   \rangle 
=  \frac{1}{8} \left(\textrm{Tr}(Y) \right)^2-\frac{1}{4} \textrm{Tr}(Y^2).
\end{equation} 

The 3rd-order term has two disconnected terms as $\langle \hat{A}\hat{B}\rangle_c^3$ and $\langle \frac{\hat{A}^2}{2!} \frac{\hat{B}^2}{2!} \rangle_c \langle \hat{A}\hat{B}\rangle_c$ 
and one connected term as $\langle \frac{\hat{A}^3}{3!} \frac{\hat{B}^3}{3!}  \rangle_c$. 
One of the disconnected terms is shown by 
\begin{equation}
  \frac{1}{\left( 3!\right)^2} ({3!})
   \langle \hat{A}\hat{B}  \rangle _c ^3 
=  \frac{1}{3!} \left(\frac{\textrm{Tr}(Y^1)}{-2\cdot 1} \right)^3.
\end{equation}
The other disconnected terms are shown by 
\begin{equation}
   \langle\frac{\hat{A}^2}{2!} \frac{\hat{B}^2}{2!} \rangle_c 
   \langle  \hat{A}\hat{B}\rangle_c
= \left( \frac{\textrm{Tr}(Y^2)}{-2\cdot 2} \right)
  \left(\frac{\textrm{Tr}(Y^1)}{-2\cdot 1} \right).
\end{equation}
The connected term is shown by 
\begin{equation}   
\langle \frac{\hat{A}^3}{3!} \frac{\hat{B}^3}{3!} \rangle_{c} 
=\left( \frac{\textrm{Tr}(Y^3)}{-2\cdot 3} \right),
\label{connect3}
\end{equation}
where the coefficient $\frac{1}{-2\cdot 3}$ is explained in Appendix C.
The 3rd order term is, therefore, given by
\begin{equation}
\langle \frac{\hat{A}^3}{3!} \frac{\hat{B}^3}{3!}  \rangle 
= -\frac{1}{48} \left(\textrm{Tr}(Y) \right)^3+\frac{1}{8} \textrm{Tr}(Y)\textrm{Tr}(Y^2)  -\frac{1}{6} \textrm{Tr}(Y^3) .
\end{equation} 

In general, we consider $k$-th order term, which has several disconnected terms and one connected term. The disconnected terms are shown as 
$\langle \hat{A}\hat{B}\rangle_c^k$,
$\langle \frac{1}{2!}\hat{A}^2 \frac{1}{2!}\hat{B}^2\rangle_c \langle \hat{A}\hat{B}\rangle_c^{(k-2)}$,
$\langle \frac{1}{3!}\hat{A}^3 \frac{1}{3!}\hat{B}^3 \rangle_c \langle \hat{A}\hat{B}\rangle_c^{(k-3)}$, 
$\langle \frac{1}{4!}\hat{A}^4\frac{1}{4!}\hat{B}^4 \rangle_c \langle \hat{A}\hat{B}\rangle_c^{(k-4)}$, $\langle \frac{1}{2!}\hat{A}^2\frac{1}{2!}\hat{B}^2 \rangle_c^2 \langle \hat{A}\hat{B}\rangle_c^{(k-4)}$,  $\cdots$. 
The $p_q$ power of $q$-th connected term, $\langle\frac{1}{q!}\hat{A}^q\frac{1}{q!}\hat{B}^q \rangle_c^{p_q}$, is shown as 
\begin{equation}   
\langle \frac{\hat{A}^q}{q!} \frac{\hat{B}^q}{q!}  \rangle_{c}^{p_q} = 
\left( \frac{\textrm{Tr}(Y^q)}{-2q} \right)^{p_q}.
\end{equation}
The $k$-th term is thereby expressed as
\begin{eqnarray}
    \langle \frac{\hat{A}^k}{k!}  \frac{\hat{B}^k}{k!} \rangle &=& \sum_{p_1+2p_2+\cdot\cdot\cdot +qp_q=k} 
    \frac{1}{p_1!p_2! \cdots p_q!}\left( \frac{\textrm{Tr}(Y^1)}{-2\cdot 1} \right)^{p_1}\nonumber \\
    && \left( \frac{\textrm{Tr}(Y^2)}{-2\cdot 2} \right)^{p_2}
    \cdot \cdot \cdot 
    \left( \frac{\textrm{Tr}(Y^q)}{-2q} \right)^{p_q}.
\label{general_k}
\end{eqnarray}
In Eq.(\ref{general_k}), the connected term is shown as
\begin{equation}   
\langle \frac{\hat{A}^k}{k!} \frac{\hat{B}^k}{k!}  \rangle_{c} = \left( \frac{\textrm{Tr}(Y^k)}{-2k} \right),
\label{connected_k} 
\end{equation}
where the coefficient $\frac{1}{-2k}$ is explained in Appendix C.

\subsection{Onishi formula via the linked cluster theorem}
According to the linked cluster theorem, the logarithm of the overlap matrix element is expressed with
its connected diagrams as  
\begin{equation}
\it{ln}\langle   e^{\hat{A}} e^{\hat{B}}\rangle=\langle   e^{\hat{A}} e^{\hat{B}}\rangle_c.
\label{linked_cluster} 
\end{equation}
The contribution of connected diagrams for the overlap matrix element is given with the summation of all connected terms Eq.(\ref{connected_k}) as
\begin{equation}
\langle   e^{\hat{A}} e^{\hat{B}}\rangle_c=
\sum_{k=0}^\infty \langle \frac{\hat{A}^k}{k!} \frac{\hat{B}^k}{k!} \rangle_c=
 \sum_{k=1}^{\infty} \left( \frac{\textrm{Tr}(Y^k)}{\textrm{-2}k} \right),
\label{infinite_sum_k} 
\end{equation}
where this expression becomes an infinite series because 
the connected terms, $\langle \frac{\hat{A}^k}{k!} \frac{\hat{B}^k}{k!} \rangle_c $ are not necessarily zero for $k>N/2$ although $\langle \frac{\hat{A}^k}{k!} \frac{\hat{B}^k}{k!} \rangle $, which is a sum
of the connected terms and the disconnected terms as shown in Eq.(\ref{general_k}),
are always zero for $k>N/2$  due to Fermi statistics.
As a result, the overlap matrix element is also expressed by an infinite series through Eq.(\ref{linked_cluster}), although Eq.(\ref{AkBk}) is a finite series.
This fact means that the overlap matrix element can be expressed in two analytically different ways.

Next, we continue to investigate Eq.(\ref{infinite_sum_k}). The eigenvalues of the $Y$ matrix are denoted as $e_i$ ($i=1,\cdots,N$). Then $\textrm{Tr}(Y)=\sum_{i=1}^{N}e_i$. The summation of the connected terms is shown as  
\begin{equation}
 \langle   e^{\hat{A}} e^{\hat{B}}\rangle_c=
 \frac{1}{2} \sum_{i=1}^{N}\sum_{k=1}^{\infty}(-\frac{e_i^k}{k})
 = \frac{1}{2} \it{ln} \xi,
 \label{eigen_rep}
\end{equation}
where $\xi=\prod_{i=1}^{N}(1-e_i)$. 
Here we consider the complex logarithmic function $ln(1-z)$ by its power series $ln(1-z)=-z-\frac{z^2}{2}-\frac{z^3}{3}-\cdots $, which has the convergence radius $|z|<1$.
For the domain beyond the convergence radius, the logarithmic function can be defined by the analytic continuation, except on the singularity at $z=1$.
Note that, as for $z=1$, the logarithmic function diverges, this point cannot be defined and is a singularity. The existence of this singularity was reported as the “nodal line” (a collection of zeros of the norm overlap) through the numerical investigation \cite{OT05}, which was found to be
the major obstacles in the implementation of the Hara-Hayashi-Ring method  \cite{HHR82}.

Now we show the overlap matrix element by
\begin{equation}
\langle   e^{\hat{A}} e^{\hat{B}}\rangle 
=e^{ \langle   e^{\hat{A}} e^{\hat{B}} \rangle_c }
=e^{\frac{1}{2} \textit{ln} \xi }= \pm pv.\xi^\frac{1}{2}   
\label{sqrt_xi}
\end{equation}
where $pv.$ means principal value and square root function appears.
In the case of the inverse function of the logarithm as a complex function, that is, $e^{ln z}=z$, the infinite-multivalued nature of complex logarithm vanishes. In this case, due to the factor of $\frac{1}{2}$,
we have to discriminate the Riemann surface as $pv. \it{ln}\xi^\frac{1}{2}+ \pi m i$ with integer $m$. 
Its exponential function becomes two-valued because $e^{\pi m i}$ gives $\pm 1$ for $m$ = even or odd.
Because $ln \xi$ is the trace of  matrix $ln (I-Y)$, the overlap is shown by 
\begin{equation}
\langle   e^{\hat{A}} e^{\hat{B}}\rangle 
=e^{\frac{1}{2}\textrm{Tr}\textit{ln}(I-Y)}= \sqrt{\textrm{Det}(I-Z^{0*}Z^{1})}.
\end{equation}
This is the well-known form of the Onishi formula \cite{OY66}. Clearly, the present expression contains the square root function and it suffers from the sign problem. As discussed above, the origin of the square root is due to the infinite series expansion concerning the connected diagrams, which results in the logarithm with the $\frac{1}{2}$ factor.

Now we go back to Eq.(\ref{eigen_rep}), where we use the eigenvalues $e_i$'s of the $Y$ matrix.
As this $Y$ matrix is a product of two skew-symmetric matrices, the eigenvalues of such a matrices
are doubly-degenerated \cite{Dra52}. Therefore, we can rewrite Eq.(\ref{eigen_rep}) as,
\begin{equation}
 \langle  e^{\hat{A}} e^{\hat{B}} \rangle_c=
 \sum_{i=1}^{N/2}\sum_{k=1}^{\infty}(-\frac{e_i^k}{k})
 = \it{ln} \bar{\xi},
 \label{eigen_rep_half}
\end{equation}
where the eigenvalues of the $Y$ matrix are pair-wisely denoted as \{$e_i,e_i$\} ($i=1,\cdots,N/2$)
and $\bar{\xi}=\prod_{i=1}^{N/2}(1-e_i)$.
This double degeneracy of the eigenvalues was also rediscovered by K. Neerg\aa rd and E. W\"ust\cite{NW83}, who proved it indirectly. 
This doubly degenerate nature cancels the $\frac{1}{2}$ factor of the logarithm in Eq.(\ref{eigen_rep})
that is the direct origin of the sign problem. 
Eq. (\ref{sqrt_xi}) can be, thereby, changed to 
\begin{equation}
e^{ \langle  e^{\hat{A}} e^{\hat{B}} \rangle_c }
=e^{\it{ln} \bar{\xi} }=\bar{\xi}.
\label{without_sqrt_xi}
\end{equation}  
Therefore, the skew-symmetric property of  $Z^{0}$ and $Z^1$ matrices, which comes from Fermi statistics, can fundamentally and completely remove the sign problem from the Onishi formula.
Thus, paying attention to such a double degeneracy of the eigenvalues, we have directly derived
the sign-problem-free version of the Onishi formula from the definition of the overlap 
matrix element, Eq.(\ref{overlapzz}). 

\section{ Summary } 
We investigated two kinds of analytically different formulae 
for the overlap matrix elements between HFB wave functions.
One is the Onishi formula that was derived half a century ago \cite{OY66}.
This formula is, in general, two-valued as a complex function and has the sign problem. 
The other is Robledo's Pfaffian formula \cite{Rob09}, which has been derived recently. This formula is single-valued and is free of the sign problem.
It is theoretically interesting to investigate why there exist two analytically different formulae and why the sign problem occurs only in the Onishi formula.

To understand both formulae more deeply, we began with the Thouless representation \cite{RS80},\cite{Th60} of the HFB wave function in Eq.(\ref{hfbwave}) and the overlap matrix element in Eq.(\ref{overlapzz}).
By a naive series expansion of the exponential operators in Eq.(\ref{overlapzz}),
the overlap matrix element can be, in principle, 
expressed with a polynomial with respect to $Z$'s matrix elements due to the Fermi statistics, as shown in Eq.(\ref{AkBk_structure}). 
Thereby, the overlap is essentially single-valued although such a simple expansion does not give rise to any useful formula. 
Hence, we investigate various summation methods of the series expansion.

First, by expanding the exponential operator in the HFB wave function in Eq.(\ref{hfbwave}), 
we found that the HFB wave function can be expressed with the Pfaffians in Eq.(\ref{HFBwavePf}),
which is a finite series due to Fermi statistics. 
The overlap matrix element can be, thereby, rewritten by the product-sum form of the Pfaffians in Eq.(\ref{overlapPfPf}), which is also a finite series as the naive expansion.
Thanks to the product-sum identity of the Pfaffian Eq.(\ref{prosum}), we can sum up these Pfaffians, and finally, we succeeded in algebraically deriving Robledo's Pfaffian formula \cite{Rob09}. This derivation shows 
a relation between the finite series expansion and Robledo's Pfaffian formula \cite{Rob09}, as well as its single-valued property and the sign-problem-free nature.

Next, starting with the overlap matrix element in Eq.(\ref{overlapzz}), we evaluated the Onishi formula with the linked cluster expansion \cite{BM63}. 
We investigated the summation procedure in the series expansions of the overlap matrix element in detail, where an infinite series of the connected diagrams shows up.
This infinite summation can alter the analytical property. 
In fact, the summation for the connected diagrams leads to the logarithm with the factor $\frac{1}{2}$. As the overlap matrix element is given by the exponential function of the summation of the connected diagrams, 
this factor $\frac{1}{2}$ results in the square root function, which can be considered as the origin of the sign problem.
We also pointed out that the sign problem is completely clarified by paying attention to a mathematical fact that the eigenvalues of a product of two skew-symmetric matrices are always doubly-degenerated \cite{Dra52}. More details and further considerations about this aspect are to be discussed elsewhere  \cite{OMS17}. This double degeneracy removes the factor $\frac{1}{2}$, and the sign problem disappears.
 
Finally, through this study, we showed a case that the linked cluster theorem gives a different analytic expression for the matrix element from its original analytic property. It may be interesting to find other cases with regard to the sign problem caused by the use of the linked cluster theorem.

\section{acknowledgment}
NS acknowledges the supports by 'Priority Issue on post-K computer'
(Elucidation of the Fundamental Laws and Evolution of the Universe)
and KAKENHI grant No. 17K05433.

\appendix

\section{ Basic identities of the Pfaffian }
The Pfaffian is defined for a skew-symmetric matrix $A=(a_{ij})$ with dimension $2n\times 2n$, as   
\begin{equation}
\text{\it Pf}\left[A\right]\displaystyle \equiv\frac{1}{2^{n}n!}\sum_{\sigma\in S_{2n}}
{\rm sgn}(\sigma)\prod_{i=1}^{n}a_{\sigma(2i-1)\sigma(2i)},
\label{defpfaff}
\end{equation}
where $S_{2n}$ is the symmetry group of degree ($2n$). The $\sigma$ is a permutation of $\{1,2,3,\cdots , 2n\}$ and  ${\rm sgn}(\sigma)$ is its sign.
If we impose a condition,$\sigma(2i-1)<\sigma(2i)$ for $1\le i \le n$ on $\sigma$, 
$2^{n}$ duplications for $\sigma$ are the same contribution in 
Eq.(\ref{defpfaff}) and we can remove $2^{n}$ multiplicity. 
Furthermore, if we impose a condition as 
\begin{eqnarray}
&&\sigma(2i-1)<\sigma(2i)\,\,\, \text{for} \,\,\,  1\le i \le n   \nonumber \\ 
&&\sigma(1)<\sigma(3)<\cdots < \sigma(2n-1),
\label{pfcond}
\end{eqnarray}
$2^{n}n!$ duplications give also the same contribution  in Eq.(\ref{defpfaff}).
The definition of the Pfaffian can be, therefore, rewritten as  
\begin{equation}
\text{\it Pf}\left[A\right]\displaystyle \equiv\sum_{\sigma}
{\rm sgn}(\sigma)\prod_{i=1}^{n}a_{\sigma(2i-1)\sigma(2i)},
\label{defpfaff1}
\end{equation}
where $\sigma$ is taken under the above restriction, Eq.(\ref{pfcond}).

For an $n \times n$ $(n=\text{odd})$ skew-symmetric matrix,
$\text{\it Pf}\left[A\right]  =0$. For a $2 \times 2$ skew-symmetric matrix,  $\text{\it Pf}\left[A\right] =a_{12}$.
For a $4 \times 4$ skew-symmetric matrix, 
\begin{equation}
\text{\it Pf}\left[A\right] =a_{12}a_{34}-a_{13}a_{24}+a_{14}a_{23}.
\label{pfaff4}
\end{equation} 
For a skew-symmetric matrix $A$ with dimension $2n \times 2n$, the following relations hold as
\begin{equation}
{\it Pf} \left[ -A \right   ]= (-)^n {\it Pf} \left[A\right   ],
\label{minuspf}
\end{equation} 
and
\begin{equation}
{\it Pf} \left[Q^T A Q\right   ]=\textrm{ Det} \left[Q\right] {\it Pf} \left[A\right   ],
\label{detpf}
\end{equation} 
where $Q$ is an arbitrary $2n \times 2n$ matrix.

The Pfaffian of skew-symmetric block diagonal matrix $A$ with dimension $2n \times 2n$ becomes a product of 
the Pfaffians of $n \times n$ sub-matrices $A_1$ and $A_2$ as 
\begin{equation}
{\it Pf}
 \left[ \begin{array}{cc}
                A_1    &    0      \\
                0   &      A_2         
             \end{array} \right] = {\it Pf} \left[ A_1 \right] \cdot {\it Pf}  \left[ A_2 \right] .
\label{pf12}
\end{equation} 
For a special block skew-symmetric matrix with $n \times n$ sub-matrices $C$,  the following identity holds as,
\begin{equation}
{\it Pf}
 \left[ \begin{array}{cc}
                0    &    C      \\
               -C^T   &      0         
             \end{array}\right   ]
   =(-1)^{n(n-1)/2}\textrm{ Det} \left[  C\right   ].
\label{CCpf}
\end{equation}

\section{ Product-sum identity of the Pfaffian }

For arbitrary skew-symmetric matrices $A$ and $B$, the following product-sum identity holds as
\begin{equation}
{\it Pf}\left [A+B\right]=\sum_{r=0}^{m/2}\sum_{{\mathcal I} \in I^m_{2r}}(-1)^{|{\mathcal I}|-r}{\it Pf}\left [A_{\mathcal I}\right]
{\it Pf}\left [B_{\overline{{\mathcal I}}}\right]
\label{prosum}
\end{equation}
where ${\mathcal I}=\{ i_1, i_2,  \cdots, i_{2r} \} $ is a subset with $2r$ elements of $\{ 1,2, \cdots, m \}$ and $||$ means sum of the elements, that is,
$|{\mathcal I} |=i_1+i_2+ \cdots +i_{2r}$.
$\overline{{\mathcal I}}$ is the complementary set of ${\mathcal I}$ concerning $\{ 1,2, \cdots, m \}$.
This proof is given by papers in pure mathematics \cite{IW99} and \cite{ST90}. 

\section{ Derivation of connected term} 
 
Here we briefly discuss how to derive Eqs.(\ref{connect2},\ref{connect3},\ref{connected_k}).
Let us begin with the $k$-th order matrix element for connected diagram given by 
\begin{eqnarray}
  & &  \langle \frac{\hat{A}^k}{k!}  \frac{\hat{B}^k}{k!} \rangle_c   \nonumber \\
  &=& \frac{1}{\left(k!\right)^2}  \frac{1}{2^{2k}} 
  \sum_{p,q,p',q'}Z^{0*}_{p'_1,q'_1}Z^{1}_{p_1,q_1}  \cdots {Z^{0*}}_{p'_k,q'_k}{Z^{1}}_{p_k,q_k} \nonumber \\
  & & \langle - | c_{q'_k}  c_{p'_k} \cdots c_{q'_1}  c_{p'_1}  
   c^{\dagger}_{p_1} c^{\dagger}_{q_1} \cdots  c^{\dagger}_{p_k} c^{\dagger}_{q_k} |- \rangle_c.
 \label{AkBkZ}  
\end{eqnarray}

To evaluate this matrix element, we consider it diagrammatically.
Figs. 1(a) and 1(b) show the pair-annihilation operator and the pair-creation operator, respectively.
Fig. 1(c) illustrates Eq.(\ref{AkBkZ}), where
the $Z^{0*}$ and $Z^{1}$ matrices are alternately linked and are shown by a ring with bipartite structure graphically. 

For example, we consider one of contractions as
\begin{equation}
\langle c_{q'_1}c_{p_1}^{\dagger} \rangle \langle c_{p'_2}c_{q_1}^{\dagger} \rangle 
 \langle c_{q'_2}c_{p_2}^{\dagger} \rangle \langle c_{p'_3}c_{q_2}^{\dagger} \rangle \cdots
 \langle c_{q'_k}c_{p_k}^{\dagger} \rangle \langle c_{p'_1}c_{q_k}^{\dagger}\rangle,
\end{equation}
which gives $-\delta_{q'_1,p_1}\delta_{q_1,p'_2}\delta_{q'_2,p_2} \cdots \delta_{q'_k,p_k}\delta_{q_k,p'_1}$. The corresponding matrix element of Eq.(\ref{AkBkZ}) 
becomes $-\textrm{Tr}(Y^k)$.
As the degeneracy concerning the interchange of the indices in Eq.(\ref{AkBkZ}) is 
$2^{2k-1} (k!)^2/k$, the $k$-th order connected overlap matrix element becomes $-\frac{1}{2k}\textrm{Tr}(Y^k) $. 

\begin{figure}[htb]
\includegraphics[width=7cm]{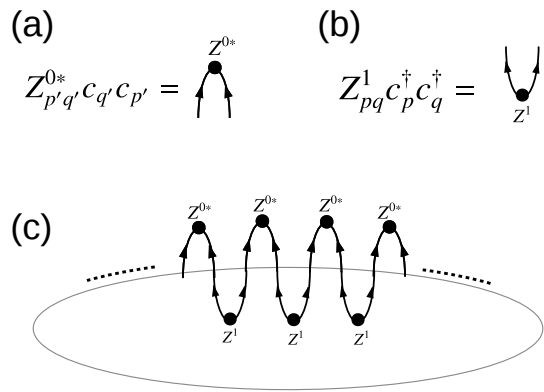}
\caption{(a) Pair-annihilation operator, (b) Pair-creation operator, (c) Ring structure of connected diagram.}
\label{connected_ring}
\end{figure}


\end{document}